\documentclass[%
 preprint, titlepage,
 amsmath,amssymb,
 aps,
]{revtex4-2}
\usepackage{amsthm}
\usepackage[utf8]{inputenc}
\usepackage{url}

\newtheorem{proposition}{Proposition}

\newtheorem{definition}{Definition}

\newcommand{\bA}{\boldsymbol{A}}
\newcommand{\bB}{\boldsymbol{B}}
\newcommand{\bC}{\boldsymbol{C}}
\newcommand{\bD}{\boldsymbol{D}}
\newcommand{\ba}{\boldsymbol{a}}
\newcommand{\bb}{\boldsymbol{b}}
\newcommand{\bc}{\boldsymbol{c}}
\newcommand{\bu}{\boldsymbol{u}}
\newcommand{\bx}{\boldsymbol{x}}
\newcommand{\by}{\boldsymbol{y}}
\newcommand{\bz}{\boldsymbol{z}}
\newcommand{\bomega}{\boldsymbol{\omega}}
\newcommand{\balpha}{\boldsymbol{\alpha}}
\newcommand{\bbeta}{\boldsymbol{\beta}}
\newcommand{\bgamma}{\boldsymbol{\gamma}}
\newcommand{\bnabla}{\boldsymbol{\nabla}}
\newcommand{\bcdot}{\boldsymbol{\cdot}}
\newcommand{\upi}{\pi}

\begin{document}
\title{New relations for energy flow in terms of vorticity}

\author{Paul Valiant} \email{pvaliant@gmail.com}
\affiliation{Department of Computer Science, Brown University, Providence RI 02912, USA}
\date{\today}

\begin{abstract}
    Considering the vorticity formulation of the Euler equations, we partition the kinetic energy into its contribution from each pair of interacting vortices. We call this contribution the ``interaction energy". We show that each contribution satisfies a reciprocity relation on triples of vortices: $\bA$'s action on $\bB$ changes the interaction energy between $\bB$ and $\bC$ in an equal and opposite way to the effect of $\bC$'s action on $\bB$ on the interaction energy between $\bA$ and $\bB$. This result is a curiously detailed accounting of energy flow, as contrasted to standard pointwise conservation laws in fluid dynamics. This result holds for all triples of points $\bA,\bB,\bC$ in two dimensions; and in 3 dimensions for all points $\bA,\bC$, and all closed vorticity streamlines $\bB$. We show this result in 3 dimensions as a consequence of an interaction energy flow around $\bB$ that is a function only of the triple $(\bA,\bb\in \bB,\bC)$, a result which may be of independent interest.
\end{abstract}
\maketitle
\section{Introduction}
One of the principal challenges of understanding the global behavior of evolving fluids is that the known conservation laws for the Euler (and Navier-Stokes) equations provide only weak control over such evolution. The standard conserved quantities for the Euler equation are the three standard physics measures of momentum $\int \bu\,d\bx$, angular momentum $\int \bx\times \bu\,d\bx$, and energy $\frac{1}{2}\int ||\bu||^2\,d\bx$, along with three more specialized quantities, the helicity $\int \bu\bcdot\bomega\,d\bx$, the fluid impulse $\frac{1}{2}\int \bx\times\bomega\,d\bx$, and the moment of the fluid impulse $\frac{1}{3}\int \bx\times (\bx\times \bomega)\,d\bx$~\citep{Majda-Bertozzi}. In addition, the Euler equations explicitly describe how velocity is advected (up to the action of the gradient of pressure), and the vorticity formulation of the Euler equations describes how vorticity is advected with the fluid. These observations constitute pointwise conservation laws, from which one can easily derive the conservation of circulation along any closed loop transported with the fluid~\citep{Tao-blog-conserved}. 
Of these invariants, energy is arguably the most important for the field of PDEs, its nonnegativity being the source of many inequalities and mono-variants, which collectively are described as ``energy methods"~\citep{beale1984remarks, he2005regularity, hou2005global,vallis1989extremal,davidson1994global,peng2008convergence} (and see \citep{tao2006nonlinear} for a textbook treatment). Understanding how energy flows between different interacting modes of a fluid has also underpinned much of the study of turbulence since Kolmogorov's introduction of the ``energy cascade"~\citep{kolmogorov,pope2001turbulent}.



Here we introduce a new relation describing how energy is partitioned between different interactions in the fluid. The main result, Proposition~\ref{Proposition:3d}, describes essentially, how for a triple of regions of the fluid $\bA,\bB,\bC$, the effect $\bA$ has on the energy between $\bB$ and $\bC$ is exactly canceled out by the effect $\bC$ has on the energy between $\bB$ and $\bA$ (where the energy ``between" two regions of fluid will be introduced shortly). In contrast to much previous literature on conservation laws involving the whole fluid, or an advecting closed loop (in the case of conservation of circulation), our relation holds for triples of locations, and thus provides a rather more detailed accounting of the flow of energy in an evolving fluid.

The results here are phrased in terms of a fairly straightforward vorticity-based view of energy, though one that does not seem to be highlighted in the literature: the Biot-Savart law expresses velocity as a linear function of vorticity, and since energy is the 2-norm of velocity, we have that energy is a bilinear function of vorticity; the bilinear contribution to energy induced by the pair of regions $\bA$ and $\bB$ (of the the vorticity field) we denote as the ``interaction energy" between these regions. 
Explicitly, we define the ``interaction energy'' between any two vortices to be the contribution that the pointwise product of their respective induced velocities makes to energy (see Section~\ref{sec:interaction} for details). Intuitively, two nearby parallel vortices have a positive interaction energy, while two nearby antiparallel vortices have a negative interaction energy, and the effect is stronger when the displacement between two parallel vortices is along the axis of their vorticity as opposed to orthogonally. This view encourages us to ask local questions: if a vortex gets advected closer to another vortex, where does this energy come from? The answer, as described above, is a local reciprocity relation: if $\bA$ acts on $\bB$ to increase the interaction energy between $\bB$ and $\bC$, then this effect is exactly counteracted by the effect $\bC$'s advection of $\bB$ has on the interaction energy between $\bA$ and $\bB$.

For fluids in two dimensions, this result holds when $\bA,\bB,\bC$ are arbitrary points in the fluid, but in three dimensions we must restrict $\bB$ to be a complete streamline of the vorticity (also known as a \emph{vortex filament}). In this sense, the result loses ``resolution" in three dimensions, since it does not describe the energy balance for individual points $\bb\in \bB$, but only aggregated over the entirety of a streamline $\bB$.

To shed light on this, our final result (Proposition~\ref{Proposition:flow}) expresses the flow of the above energy discrepancy around a vortex streamline $\bB$ as the spatial derivative of an energy ``flux" for the triple $\bA,\bb\in\bB,\bC$ that depends only on the points $\bA,\bb\in \bB,\bC$, with no dependence on the rest of the vortex streamline $\bB$. 

We note that all our results are expressed in terms of points in the vorticity field, but since interaction energy is bilinear, these results (including the definition of interaction energy, and the various conservation relations) may easily be integrated over regions of space to define interaction energy relations between \emph{regions} of the fluid.

For previous work on interacting triplets in fluid dynamics though in a rather different setting of Fourier modes in turbulent flow, see~\citep{moffatt}.
\subsection{Perspective}

Many of the proposals for potential blowup solutions of the Euler and Navier-Stokes equations involve a rapid local accumulation of energy~\citep{moffatt2019towards,pelz1997locally,kerr2013growth}. 
The results of this paper impose new conditions on local energy transfer in a fluid, where, intuitively, if two vortices approach each other and accumulate energy, this must be because a third vortex pushed them together, and this third vortex must lose energy as a result of being pushed away (or twisted) by the first two vortices. While such a characterization is hard to formalize, this intuition that blowup must be fueled by the blowup region continually ejecting vorticity comparable to what it contains, argues against a blowup configuration with constant circulation about the blowup point. We thus hope that these results might shed light on some basic structural questions of the evolution of the Euler and Navier-Stokes equations.

\section{Interaction energy}\label{sec:interaction}

We consider the Euler equation, expressed in terms of the vorticity $\bomega\triangleq \bnabla\times \bu$ where $\bu$ is the velocity field of the fluid. The evolution of $\bomega$ is described by \[\frac{d\bomega}{dt}=(\bomega\bcdot\bnabla)\bu-(\bu\bcdot\bnabla)\bomega\]
where the velocity is implicitly defined from vorticity via the Biot-Savart law, $\bu(\bx)=\int \frac{1}{4\upi}\,\frac{\bomega(\by)\times (\bx-\by)}{||\bx-\by||^3}\,d\by$.

We begin by reformulating energy into an equivalent expression, expressed as a double integral over pairs of points in the vorticity  of the fluid, of a quantity we term the ``interaction energy" of the vorticity at those two points. Explicitly, the energy of a fluid with velocity $\bu(\bx)$ is $E=\frac{1}{2}\int \bu(\bx)\bcdot \bu(\bx)\,d\bx$, which we reexpress in terms of the vorticity $\bomega(\bx)=\bnabla\times \bu(\bx)$ and the Biot-Savart law. Expanding each copy of $\bu$ in the definition of energy, and moving the $\bx$-integration to the inside yields \[E=\frac{1}{32\upi^2}\int\int\int\frac{(\bomega(\by)\times (\bx-\by))\bcdot (\bomega(\bz)\times(\bx-\bz))}{||\bx-\by||^3\, ||\bx-\bz||^3}\,d\bx\,d\by\,d\bz\]
We simplify this expression by expressing the innermost  integral (over $\bx$) as a function of the three vectors $\bomega(\by)$, $\bomega(\bz)$, and $\bz-\by$ (since the $\bx$-integral is invariant to translations of \emph{both} $\by$ and $\bz$, it depends only on the difference $\bz-\by$). Define the \emph{interaction energy} to be \[I(\bomega(\by),\bomega(\bz),\bz-\by)\triangleq\frac{1}{32\upi^2}\int \frac{(\bomega(\by)\times (\bx-\by))\bcdot (\bomega(\bz)\times(\bx-\bz))}{||\bx-\by||^3\, ||\bx-\bz||^3}\,d\bx\]

A straightforward computation (see Appendix~\ref{sec:interaction-energy-derivation}) yields the following direct formula for interaction energy:
\[I(\bomega(\by),\bomega(\bz),\bz-\by)=\frac{1}{8\upi ||\bz-\by||} \left[\bomega(\by)_{\parallel \bz-\by}\bcdot\bomega(\bz)_{\parallel \bz-\by}+\frac{1}{2}\bomega(\by)_{\bot \bz-\by}\bcdot\bomega(\bz)_{\bot \bz-\by}   \right],\]
where we use the subscripts ${}_{\parallel \bz-\by}$ and ${}_{\bot \bz-\by}$ to denote projections of vectors into the components parallel to $\bz-\by$ and perpendicular to $\bz-\by$ respectively. Intuitively, two vortices have an energy interaction if they are parallel; the strength of the interaction varies inversely with the distance and proportionally to the two vorticities; and the interaction is twice as strong when the displacement is in the direction of both vorticities, as opposed to in a transverse direction.

Thus the energy of a fluid equals $E=\int\int I(\bomega(\by),\bomega(\bz),\bz-\by)\,d\by\,d\bz$, and we have reexpressed energy in a more ``microscopic" sense, revealing how interactions between vorticity at different locations comprise different components of the overall energy.

We point out, lest it be overlooked, that the interaction energy between two particular vortices \emph{may} be negative (if, for example, the vortices are anti-parallel); though of course the double integral of interaction energy over all pairs of vortices equals the standard energy and is nonnegative.
\section{Detailed local conservation of interaction energy}

\subsection{The 2-dimensional setting}
To motivate what follows, we consider the 2-dimensional case. Since vorticity is always ``out of the plane", the expression for 2d interaction energy is much simpler\footnote{The derivation is analogous to the 3d case, though with the caveat that fluid flows in 2d have \emph{infinite} energy unless their total vorticity is 0 (or different boundary conditions can be imposed). The 2d interaction energy here ignores this divergence at infinite distances.}: $I_{2d}(\bomega(\by),\bomega(\bz),\bz-\by)=\frac{1}{4\upi} \bomega(\by)\bomega(\bz) \log|\bz-\by|$. We now examine how interaction energy evolves in time under the Euler equations. From the vorticity formulation of the Euler equations, we view each vortex as contributing to the velocity at all other locations, via the Biot-Savart law, where vortices will be advected according to their total velocity. Given 3 locations in the fluid $\bA, \bB, \bC$, we can ask how the velocity that vortex $\bA$ induces at location $\bB$ advects $\bB$ so as to change its interaction energy with vortex $\bC$. Explicitly, this effect is proportional to the amount that the velocity field induced by vortex $\bA$ moves $\bB$ towards $\bC$, which is easily calculated as $\frac{\bomega(\bA)}{2\upi}\,\frac{(\bB-\bC)\times (\bB-\bA)}{ ||\bB-\bA||^2\,||\bB-\bC||}$. To calculate the effect this motion has on the interaction energy, we use the chain rule to see that the derivative of the log of the distance $||\bB-\bC||$ is the derivative of the distance (which we just computed) times $\frac{1}{||\bB-\bC||}$; and we multiply by $\frac{1}{4\upi}\bomega(\bB)\bomega(\bC)$ according to the interaction energy formula to yield a change in interaction energy of \[\frac{\bomega(\bA)\bomega(\bB)\bomega(\bC)}{8\upi^2} \frac{(\bB-\bC)\times (\bB-\bA)}{ ||\bB-\bA||^2\,||\bB-\bC||^2}\]

The crucial observation is that this quantity is antisymmetric with respect to swapping $\bA$ and $\bC$, leading to the following result:

\begin{proposition}\label{Proposition:2d}
In 2 dimensions, for any three points $\bA,\bB,\bC$, in coordinates advecting with the fluid, $\bA$'s action on $\bB$ affects the interaction energy between $\bB$ and $\bC$ in an exactly equal and opposite way as $\bC$'s action on $\bB$ affects the interaction energy between $\bA$ and $\bB$.
\end{proposition}

\subsection{The 3-dimensional setting}

Moving to 3 dimensions introduces many new effects, including vortex stretching as a result of advection. The exact analogue to Proposition~\ref{Proposition:2d} does not hold, but must be modified so that $\bB$ is no longer a single point, but rather a complete streamline of the vortex field (often called a \emph{vortex filament} or \emph{vortex loop}).

\begin{proposition}\label{Proposition:3d}
In 3 dimensions, for any points $\bA,\bC$ and any vortex filament $\bB$, in coordinates advecting with the fluid, $\bA$'s action on $\bB$ affects the interaction energy between $\bB$ and $\bC$ in an exactly equal and opposite way as $\bC$'s action on $\bB$ affects the interaction energy between $\bA$ and $\bB$.
\end{proposition}

As a motivating example to demonstrate that this modification is needed, consider $\bA$ to be an axisymmetric vortex loop, (or a point on such a loop) with vorticity pointing along the loop (in the transverse, $\boldsymbol{\hat{\theta}}$, direction in cylindrical coordinates), while $\bB$ and $\bC$ are points on the axis with vorticity pointing along the axis. The velocity induced by $\bA$ will push $\bB$ up or down the axis, affecting its interaction energy with $\bC$; yet $\bC$ will have no analogous effect on $\bB$'s interaction with $\bA$, since $\bC$ has axial vorticity and induces a velocity field that will leave $\bB$ exactly unchanged on the axis. Thus $\bA$ might significantly affect $\bB$'s interaction energy with a third point $\bC$; however, it turns out that if we account for $\bB$'s entire streamline, then the axial ``bunching" induced by vortices such as $\bA$ will not affect interaction energy with the entire streamline, merely move energy up and down the streamline.

While we could prove Proposition~\ref{Proposition:3d} directly, it is perhaps simpler to derive this global-over-$\bB$ conservation law by expressing it as the derivative of a certain flow along $\bB$.

\begin{definition}\label{def:flow}
Define the ``interaction energy flow" to be \begin{align*}F&_{\bomega}(\bA,\bB,\bC)\\&=\frac{1}{64\upi^2}\left[\frac{((\bA-\bB)\times\bomega(A))\bcdot\bomega(\bC)}{||\bA-\bB||^3\,||\bC-\bB||}+\frac{((\bB-\bA)\times(\bB-\bC))\bcdot\bomega(\bA)((\bB-\bC)\bcdot\bomega(\bC))}{||\bA-\bB||^3\,||\bC-\bB||^3}\right]\end{align*}
\end{definition}

\begin{proposition}\label{Proposition:flow}
In 3 dimensions, for any points $\ba,\bb,\bc$ the effect of $\ba$'s induced velocity on $\bb$ of the interaction energy between $\bb$ and $\bc$ plus, symmetrically, the effect of $\bc$'s induced velocity on $\bb$ of the interaction energy between $\ba$ and $\bb$ equals the derivative with respect to $\bb$ in the direction $\bomega(\bb)$ of the (symmetrized) interaction energy flow $F_{\bomega}(\ba,\bb,\bc)+F_{\bomega}(\bc,\bb,\ba)$.
\end{proposition}

As a corollary, if $\bB$ is a complete streamline of vorticity (that is, either a finite loop, or we assume $\bomega$ decays sufficiently at infinity), then the integral over all $\bb\in \bB$ of the derivative in the direction $\bomega(\bb)$ tangent to $\bB$ of the symmetrized energy flow $F_{\bomega}(\ba,\bb,\bc)+F_{\bomega}(\bc,\bb,\ba)$ must be 0, and thus we conclude Proposition~\ref{Proposition:3d}.

\medskip
We make two final notes. First, the notion of interaction energy extends to the Navier-Stokes equations. As the Navier-Stokes equations in their vorticity formulation describe the fluid's evolution as exactly that of the Euler equations, plus a diffusion term $\nu \boldsymbol{\Delta}\bomega$, the interaction energy evolves under the constraints of Proposition~\ref{Proposition:3d} though with an additional diffusion of interaction energy. Perhaps frustratingly, while (total) energy decreases monotonically under the Navier-Stokes equations, interaction energy does not---since, as pointed out above, interaction energy \emph{can} be negative, and thus would decay to 0 by \emph{increasing}.

Second, it is perhaps tempting to conjecture that the interaction energy of a vortex tube $\bB$ with the entire fluid $\mathbb{R}^3$ might be a conserved quantity as the fluid evolves, since Proposition~\ref{Proposition:3d} says that the effect of the entire fluid on $\bB$ changes $\bB$'s interaction energy with the entire fluid by exactly the negative of this exact same quantity, implying that this induced change in energy must equal 0. However, this omits the effect advection by $\mathbb{R}^3$ on $\mathbb{R}^3\setminus \bB$ has on this interaction energy. Such potential conservation laws, as far as we understand, do not hold.




\bibliography{bib}

\appendix
\section{Derivation of interaction energy}\label{sec:interaction-energy-derivation}
We change variables in the definition of interaction energy to avoid confusion in the calculations: \begin{equation}\label{eq:I-def-appendix}I(\bomega(\bb),\bomega(\bc),\bc-\bb)\triangleq\frac{1}{32\upi^2}\int \frac{(\bomega(\bb)\times (\ba-\bb))\bcdot (\bomega(\bc)\times(\ba-\bc))}{||\ba-\bb||^3\, ||\ba-\bc||^3}\,d\ba\end{equation}

Since this definition is rotationally symmetric, without loss of generality we consider $\bc-\bb$ to point along the axis in cylindrical coordinates, with $\bb$ being at the origin and $\bc$ being 1 unit along the axis, i.e. the unit vector $\hat{\bz}$. Since, for general vectors, $(\bA\times \bB)\bcdot (\bC\times \bD)=(\bA\bcdot \bC)(\bB\bcdot \bD)-(\bB\bcdot \bC)(\bA\bcdot \bD)$, we may simplify the numerator of Equation~\ref{eq:I-def-appendix} as \begin{equation}\label{eq:I-num}(\bomega(\bb)\times (\ba-\bb))\bcdot (\bomega(\bc)\times(\ba-\bc))=\bomega(\bb)\bcdot\bomega(\bc)(\ba\bcdot (\ba-\hat{\bz}))-(\ba\bcdot\bomega(\bc))((\ba-\hat{\bz})\bcdot\bomega(\bb))\end{equation} Since $I$ is bilinear in the inputs $\bomega(\bb),\bomega(\bc)$, it suffices to evaluate $I$ when $\bomega(\bb),\bomega(\bc)$ are pairs of basis vectors. When $\bomega(\bb),\bomega(\bc)$ are orthogonal, the first term in the right hand side of Equation~\ref{eq:I-num} equals 0. When $\bomega(\bc)=\hat{\bx}$ and $\bomega(\bb)$ has no $x$-component, then the dot product $\ba\bcdot \bomega(\bc)$ is proportional to the $x$-component of $\ba$ while the other dot product $(\ba-\hat{\bz})\bcdot\bomega(\bb)$ does not depend on the $x$-component of $\ba$, and thus the integral over all $\ba$ of this expression is 0 (as the positive-$x$ contribution is canceled out by the negative-$x$ contribution). Since $I$ is symmetric with respect to swapping its first two inputs, we generalize the previous observation to see that $I$ is 0 whenever its first two inputs are orthogonal. By the bilinearity of $I$, we need only make 2 additional calculations: when $\bomega(\bb)=\bomega(\bc)=\hat{\bx}$, and when $\bomega(\bb)=\bomega(\bc)=\hat{\bz}$.

In the first case, we have $I(\hat{\bx},\hat{\bx},\hat{\bz})=\frac{1}{32\upi^2} \int \frac{||\ba||^2 -\ba_z - \ba_x^2}{||\ba||^3\,||\ba-\hat{\bz}||^3}\,d\ba=\frac{1}{32\upi^2} \int \frac{\ba_y^2+(\ba_z^2-\ba_z)}{||\ba||^3\,||\ba-\hat{\bz}||^3}\,d\ba=\frac{1}{16\upi}$, where it is easy to verify (e.g., by integrating in Mathematica in cylindrical coordinates) that the $(\ba_z^2-\bz)$ term in the numerator integrates to 0 and it is the $\ba_y$ term that contributes $\frac{1}{16\upi}$.

In the second case we  have $I(\hat{\bz},\hat{\bz},\hat{\bz})=\frac{1}{32\upi^2} \int \frac{||\ba||^2 -\ba_z - \ba_z^2+\ba_z}{||\ba||^3\,||\ba-\hat{\bz}||^3}\,d\ba=\frac{1}{32\upi^2} \int \frac{\ba_x^2+\ba_y^2}{||\ba||^3\,||\ba-\hat{\bz}||^3}\,d\ba=\frac{1}{8\upi}$, which follows from the previous result, as the integral is symmetric with respect to swapping $x$ and $y$ coordinates.

Thus, by the rotational symmetry and bilinearity of $I$, we conclude with a complete expression for $I$, taking into account the fact that $I$ should vary inversely with its third argument, since the integral has 2 spacial terms in the numerator, 6 in the denominator, and is over 3 spacial dimensions yielding a spacial dependence of order $2+3-6=-1$: \[I(\bomega(\bb),\bomega(\bc),\bc-\bb)=\frac{1}{8\upi ||\bc-\bb||} \left[\bomega(\bb)_{\parallel \bc-\bb}\bcdot\bomega(\bc)_{\parallel \bc-\bb}+\frac{1}{2}\bomega(\bb)_{\bot \bc-\bb}\bcdot\bomega(\bc)_{\bot \bc-\bb}   \right],\]
where we use the subscripts ${}_{\parallel \bc-\bb}$ and ${}_{\bot \bc-\bb}$ to denote projections of vectors into the components parallel to $\bc-\bb$ and perpendicular to $\bc-\bb$ respectively.

\section{Proof of Proposition~\ref{Proposition:flow}}

\begin{proof}
We reexpress the projection operators in the definition of interaction energy in a more convenient form by explicitly computing the projection in terms of dot products: \begin{align}I(\bomega(\bb),\bomega(\bc)&,\bc-\bb)=\frac{1}{8\upi ||\bc-\bb||} \left[\frac{1}{2}\bomega(B)_{\bot \bc-\bb}\bcdot\bomega(\bc)_{\bot \bc-\bb} +\bomega(\bb)_{\parallel \bc-\bb}\bcdot\bomega(\bc)_{\parallel \bc-\bb}  \right]\nonumber\\\label{eq:interaction-appendix}
&=\frac{1}{16\upi ||\bc-\bb||} \left[\bomega(\bb)\bcdot\bomega(\bc)+\frac{1}{||\bc-\bb||^2}(\bomega(\bb)\bcdot(\bc-\bb))(\bomega(\bc)\bcdot(\bc-\bb))   \right]\end{align}

Advection by $\ba$ will induce a velocity at location $\bb$ of $\frac{1}{4\upi} \frac{\bomega(\ba)\times(\bb-\ba)}{||\ba-\bb||^3}$, and further, the vorticity formulation of the Euler equations says that this velocity field induces a change in vorticity at $\bb$ (in coordinates advected with the fluid) equal to its directional derivative in the direction $\bomega(\bb)$, namely, $\frac{1}{4\upi} \bomega(\bb)\bcdot \bnabla_{\bb} \frac{\bomega(\ba)\times(\bb-\ba)}{||\ba-\bb||^3}$. Proposition~\ref{Proposition:flow} states that the time derivative of interaction energy $I(\bomega(\bb),\bomega(\bc),\bc-\bb)$, when $\bb$ and $\bomega(\bb)$ have time derivatives as just derived respectively, and when added to the symmetric analogue with $\ba,\bc$ swapped, exactly equals the (spatial) derivative in the direction $\bomega(\bb)$ of the interaction energy flow of Definition~\ref{def:flow} and summed with its symmetric analogue.

We prove this in 4 parts, by separately analyzing the contributions from time derivatives of $\bb$ and $\bomega(\bb)$, on each of the two terms in the definition of interaction energy in Equation~\ref{eq:interaction-appendix}.

The contribution of the time derivative of $\bomega(\bb)$ to the first term is simply $\frac{1}{16\upi||\bc-\bb||} \bomega(\bb)'\bcdot \bomega(\bc)=\frac{1}{16\upi||\bc-\bb||} \left[\frac{1}{4\upi} \bomega(\bb)\bcdot \bnabla_{\bb} \frac{\bomega(\ba)\times(\bb-\ba)}{||\ba-\bb||^3}\right]\bcdot \bomega(\bc)$. This expression equals the directional derivative in the direction $\bomega(\bb)$ of \emph{most} of the first term of the interaction energy flow: the denominator $||\bc-\bb||$ is outside of the derivative, meaning that it remains for us to account for the derivative with respect to the remaining portion of the first term by other means.

Explicitly, the remaining portion of the directional derivative of the first term of interaction energy flow equals \begin{align}\label{eq:cancel}\frac{1}{64\upi^2}\hspace{1cm}&\hspace{-1cm}\frac{((\ba-\bb)\times\bomega(\ba))\bcdot\bomega(\bc)}{||\ba-\bb||^3}\left[\bomega(\bb)\bcdot \bnabla\frac{1}{||\bc-\bb||}\right]\nonumber\\&=\frac{1}{64\upi^2}\frac{((\ba-\bb)\times\bomega(\ba))\bcdot\bomega(\bc)((\bc-\bb)\bcdot\bomega(\bb))}{||\ba-\bb||^3\,||\bc-\bb||^3}\end{align}

We relate this to the contribution from the $\bb$ derivative of the first term of Equation~\ref{eq:interaction-appendix}, namely $\bnabla_{\bb} \frac{1}{16\upi||\bc-\bb||}\bomega(\bb)\bcdot\bomega(\bc)=\frac{1}{16\upi}\frac{\bc-\bb}{||\bc-\bb||^3}\bomega(\bb)\bcdot\bomega(\bc)$, which, after taking the dot product with the time derivative of $\bb$ given by the Biot-Savart law yields $\frac{1}{64\upi^2}\frac{(((\ba-\bb)\times\bomega(\ba))\bcdot(\bc-\bb)}{||\ba-\bb||^3\,||\bc-\bb||^3}\bomega(\bb)\bcdot\bomega(\bc)$. We claim that this equation plus its symmetric analogue exactly equals Equation~\ref{eq:cancel} plus its symmetric analogue. To show this, we consider the numerators only, each of which is a 5-way product. To simplify notation, let $\bA=\ba-\bb, \bC=\bc-\bb,\balpha=\bomega(\ba),\bbeta=\bomega(\bb),\bgamma=\bomega(\bc)$. The claim is that, for arbitrary $\bA,\bC,\balpha,\bbeta,\bgamma$ we have \begin{equation}\label{eq:identity}(\bA\times\balpha)\bcdot\bgamma (\bC\bcdot\bbeta)+(\bC\times\bgamma)\bcdot\balpha (\bA\bcdot\bbeta)=(\bA\times \balpha)\bcdot \bC (\bgamma\bcdot\bbeta)+(\bC\times \bgamma)\bcdot \bA (\balpha\bcdot\bbeta),\end{equation} where the left hand side is the sum of the numerator of Equation~\ref{eq:cancel} and its symmetric analogue, and the right hand side is the corresponding term from the interaction energy and its symmetric analogue. Equation~\ref{eq:identity} is the dot product of $\bbeta$ with a sum of 4-way products, each of which is a vector triple product multiplied by the remaining vector. Equation~\ref{eq:identity} follows from Lagrange's identity, where both sides are equal to the dot product of $\bbeta$ with the cross product $(\balpha\times\bgamma)\times(\bC\times \bA)$.

The analysis of the second term of Equation~\ref{eq:interaction-appendix} is similar, where it exactly cancels out the contributions of the second term of Definition~\ref{def:flow}. We now consider the effect of the time derivative of $\bomega(\bb)$ from the second term of Equation~\ref{eq:interaction-appendix}. As before, $\bomega(\bb)$ occurs just once in this term, in the numerator, leading to a contribution of $\frac{1}{16\upi||\bc-\bb||^3}(\bomega(\bb)'\bcdot (\bc-\bb))(\bomega(\bc)\bcdot(\bc-\bb))$, where we substitute $\bomega(\bb)'=\frac{1}{4\upi} \bomega(\bb)\bcdot \bnabla_{\bb} \frac{\bomega(\ba)\times(\bb-\ba)}{||\ba-\bb||^3}$. As above, this exactly cancels out the directional derivative with respect to $\bb$ in direction $\bomega(\bb)$ of \emph{some} of the occurrences of $\bb$ in the second term of Definition~\ref{def:flow}. What remains is:

\begin{equation}\label{eq:4th}\frac{1}{64\upi^2}\left[\bomega(\bb)\bcdot\bnabla_{\bb} \frac{(\bb-\bc)\bcdot\bomega(\bc)}{||\bc-\bb||^3}(\bb-\bc)\right]\bcdot\frac{(\ba-\bb)\times\bomega(\ba)}{||\ba-\bb||^3}\end{equation}

This expression essentially consists of a Jacobian (the entire $\bnabla_{\bb}$ term) multiplied on the left and right by vectors. What remains to show is that this expression plus its symmetric analogue equals the contribution from the $\bb$ derivative and the second term of Definition~\ref{eq:interaction-appendix}. If we consider the $\bb$ derivative applied to all of the second term \emph{except} the dot product with $\bomega(\bb)$, we have \[\left[\frac{(\ba-\bb)\times\bomega(\ba)}{4\upi||\ba-\bb||^3}\bcdot\bnabla_{\bb} \frac{1}{16\upi}\frac{(\bb-\bc)\bcdot\bomega(\bc)}{||\bc-\bb||^3}(\bb-\bc)\right]\bcdot \bomega(\bb)\]

This expression is seen to be identical to that in Equation~\ref{eq:4th} except with the Jacobian matrix multiplied on \emph{opposite} sides by the two vectors. To conclude, we calculate the Jacobian multiplied on the left by some vector $\by$ (a directional derivative in direction $\by$), yielding \begin{align}\nonumber &\by\bcdot\bnabla_{\bb} \frac{(\bb-\bc)\bcdot\bomega(\bc)}{||\bc-\bb||^3}(\bb-\bc)\\&=\frac{(\bb-\bc)\bcdot\bomega(\bc)||\bc-\bb||^2 \by+(\by\bcdot\bomega(\bc)||\bc-\bb||^2-3((\bb-\bc)\bcdot \by)((\bb-\bc)\bcdot\bomega(\bc)))(\bb-\bc)}{||\bc-\bb||^5}\label{eq:jacobian}\end{align}

Our task is to show that this expression, when substituting $\bomega(\bb)$ for $\by$ and taking a dot product with the Biot-Savart vector $\frac{\bomega(\ba)\times(\bb-\ba)}{||\ba-\bb||^3}$, is identical to swapping which vector gets substituted versus multiplied. Immediately, we see that the first term in the numerator of Equation~\ref{eq:jacobian} is in the direction $\by$, so when we take the dot product with a vector $\bz$, swapping the roles of $\by$ and $\bz$ will not change it. The third term has $\by$ appearing in a dot product with $(\bb-\bc)$, while the overall term points in the direction $(\bb-\bc)$, whose dot product is taken with $\bz$, and thus similarly, swapping the roles of $\by$ and $\bz$ leaves the third term identical. What remains is the second term in the numerator $(\by\bcdot\bomega(\bc)) (\bb-\bc)\bcdot \bz$. When substituting $\bomega(\bb)$ and the Biot-Savart vector $\frac{\bomega(\ba)\times(\bb-\ba)}{||\ba-\bb||^3}$ for $\by,\bz$ respectively, and adding in their symmetric counterparts, we obtain a multiple of the identity of Equation~\ref{eq:identity}. This concludes the proof, as all terms from both sides have exactly matched.
\end{proof}

\end{document}